\documentclass[longauth]{aa}
\usepackage{graphicx}
\usepackage{txfonts}
\usepackage{natbib}
\bibpunct{(}{)}{;}{a}{}{,} 
\usepackage[colorlinks=false]{hyperref}

\begin{document} 

   \title{Detailed stellar activity analysis and modelling of GJ 832}
   \subtitle{Reassessment of the putative habitable zone planet GJ 832c}

    \author{P. Gorrini\inst{\ref{udec}, \ref{goettingen}}
      \and N. Astudillo-Defru\inst{\ref{ucsc}}
      \and S. Dreizler \inst{\ref{goettingen}}
      \and M. Damasso \inst{\ref{inaf}}
      \and R.F. Díaz\inst{\ref{unsam}}
      \and X. Bonfils\inst{\ref{grenoble}}
      \and S.V. Jeffers \inst{\ref{mps}}
      \and J.R. Barnes \inst{\ref{openuni}}
      \and F. Del Sordo \inst{\ref{ICE}, \ref{catania}}
      \and J.-M. Almenara \inst{\ref{grenoble}}
      \and E. Artigau \inst{\ref{montreal}, \ref{omm}}
      \and F. Bouchy \inst{\ref{geneva}}
      \and D. Charbonneau \inst{\ref{harvard}}
      \and X. Delfosse \inst{\ref{grenoble}}
      \and R. Doyon \inst{\ref{montreal}, \ref{omm}}
      \and P. Figueira \inst{\ref{ESO}, \ref{porto}}
      \and T. Forveille \inst{\ref{grenoble}}
      \and C.A. Haswell \inst{\ref{openuni}}
      \and M.J. L\'opez-Gonz\'alez \inst{ \ref{granada}}
      \and C. Melo \inst{\ref{lisboa}}
      \and R.E. Mennickent \inst{\ref{udec}}
      \and G. Gaisné \inst{\ref{grenoble}}
      \and N. Morales \inst{\ref{granada}}
      \and F. Murgas \inst{\ref{iac}, \ref{laguna}}
      \and F. Pepe \inst{\ref{geneva}}
      \and E. Rodríguez \inst{\ref{granada}}
      \and N. C. Santos \inst{\ref{porto}, \ref{porto2}}
      \and L. Tal-Or \inst{\ref{israel1}, \ref{israel2}}
      \and Y. Tsapras \inst{\ref{heid}}
      \and S. Udry \inst{\ref{geneva}}
          }

   \institute{Universidad de Concepción, Departamento de Astronomía, Casilla 160-C, Concepción, Chile \label{udec}\\
              \email{paula.gorrini@uni-goettingen.de} 
              \and Institut f\"{u}r Astrophysik, Georg-August-Universit\"{a}t,  Friedrich-Hund-Platz 1, D-37077 G\"{o}ttingen, Germany\label{goettingen}
              \and Departamento de Matemática y Física Aplicadas, Universidad Católica de la Santísima Concepción, Alonso de Rivera 2850, Concepción, Chile \label{ucsc}
              \and INAF - Osservatorio Astrofisico di Torino, Via Osservatorio 20, 10025 Pino Torinese, Italy \label{inaf}
              \and International Center for Advanced Studies (ICAS) and ICIFI (CONICET), ECyT-UNSAM, Campus Miguelete, 25 de Mayo y Francia, (1650) Buenos Aires, Argentina \label{unsam}
              \and Univ. Grenoble Alpes, CNRS, IPAG, F-38000 Grenoble, France \label{grenoble}
              \and Max Planck Institute for Solar System Research, Justus-von-Liebig-weg 3, 37077 Göttingen, Germany \label{mps}
              \and School of Physical Sciences, The Open University, Walton Hall, MK7 6AA, Milton Keynes, UK \label{openuni}
              \and Institute of Space Sciences (ICE-CSIC), Campus UAB, Carrer de Can Magrans s/n, 08193, Bellaterra, Spain \label{ICE}
              \and  INAF–Osservatorio Astrofisico di Catania, via Santa Sofia, 78 Catania, Italy \label{catania}
              \and Institut de Recherche sur les Exoplanètes, Université de Montréal, Département de Physique, C.P. 6128 Succ. Centre-ville, Montréal, QC H3C 3J7, Canada \label{montreal}
              \and Observatoire du Mont-M\'egantic, Universit\'e de Montr\'eal, Montr\'eal, QC H3C 3J7, Canada\label{omm}
              \and Observatoire de Gen\`eve, Universit\'e de Gen\`eve, chemin Pegasi 51, 1290 Versoix, Switzerland \label{geneva}
              \and Center for Astrophysics | Harvard \& Smithsonian, 60 Garden Street, Cambridge, MA
              02138, USA \label{harvard}
              \and  European Southern Observatory, Vitacura, Santiago, Chile \label{ESO}
              \and Instituto de Astrofísica e Ciências do Espaço, Universidade do Porto, CAUP, Rua das Estrelas, 4150-762 Porto, Portugal \label{porto}
              \and Instituto de Astrof\'\i sica de Andaluc\'\i a (CSIC), Glorieta de la Astronom\'\i a s/n, 18008, Granada, Spain \label{granada}
              \and  Portuguese Space Agency, Estrada das Laranjeiras, n.º 205, RC, 1649-018, Lisboa, Portugal \label{lisboa}
              \and Instituto de Astrofísica de Canarias (IAC), 38205 La Laguna, Tenerife, Spain \label{iac}
              \and Departamento  de  Astrofísica,  Universidad  de  La  Laguna  (ULL),38206 La Laguna, Tenerife, Spain \label{laguna}
              \and Departamento de Física e Astronomia, Faculdade de Ciências,Universidade do Porto, Rua do Campo Alegre, 4169-007 Porto, Portugal \label{porto2}
              \and Department of Physics, Ariel University, Ariel 40700, Israel \label{israel1}
              \and Astrophysics Geophysics And Space Science Research Center, Ariel University, Ariel 40700, Israel\label{israel2}
              \and Zentrum f{\"u}r Astronomie der Universit{\"a}t Heidelberg, Astronomisches Rechen-Institut, M{\"o}nchhofstr. 12-14, 69120 Heidelberg, Germany \label{heid}
             }

   \date{Accepted May 25, 2022}

 
  \abstract
   {Gliese-832 (GJ~832) is an M2V star hosting a massive planet on a decade-long orbit, GJ~832b, discovered by radial velocity (RV). Later, a super Earth or mini-Neptune orbiting within the stellar habitable zone was reported (GJ~832c). The recently determined stellar rotation period (45.7 $\pm$ 9.3 days) is close to the orbital period of putative planet c (35.68 $\pm$ 0.03 days).
   }
   {We aim to confirm or dismiss the planetary nature of the RV signature attributed to GJ~832c, by adding 119 new RV data points, new photometric data, and an analysis of the spectroscopic stellar activity indicators.   Additionally, we  update the orbital parameters of the planetary system and search for additional signals. }
   {
   We performed a frequency content analysis of the RVs to search for periodic and stable signals. Radial velocity time series were modelled with Keplerians and Gaussian process (GP) regressions alongside activity indicators to subsequently compare them within a Bayesian framework.}
   {We updated the stellar rotational period of GJ~832 from activity indicators, obtaining 37.5 $\protect\substack{+1.4 \\ -1.5}$ days, improving the precision by a factor of 6. The new photometric data are in agreement with this value. We detected an RV signal near 18 days (FAP $<$ 4.6\%), which is half of the stellar rotation period. 
  Two Keplerians alone fail at modelling GJ~832b and a second planet with a 35-day orbital period. Moreover, the Bayesian evidence from the GP analysis of the RV data with simultaneous activity indices
  prefers a model without a second Keplerian, therefore negating the existence of planet c.
  }
  {}

   \maketitle
%

\section{Introduction}

M dwarfs are the most common stars in our Galaxy, and they are ideal targets to search for terrestrial companions due to their low masses and luminosities. Compared to the Sun, M dwarfs are smaller, and the relative decrease in their fluxes by a transiting planet of a given radius is larger (for example, by a factor of 11 for M4 dwarfs). Similarly, the reflex radial velocity (RV) amplitude of an M dwarf due to an orbiting planet of a given mass is greater (by four times for an M4 dwarf) than for a Sun-like star.
Moreover, M dwarfs have luminosities ranging from $10^{-4}$ to $10^{-1}$ $L_{\odot}$,  meaning that their habitable zones (HZs) tend to be closer than for earlier stars, typically between 0.03 and 0.4 AU \citep{Kasting}. \\
\indent A large fraction of M dwarfs are known to be magnetically active  \citep[e.g.][]{Reiners12,Jeffers18}. This activity generates  quasi-periodic RV variations that can be misinterpreted as the signature of a planetary companion \citep[e.g.][]{Queloz01,Desidera04,Bonfils07,Huelamo08,Santos14,Robertson2014b, Robertson2015}.
Typical manifestations of stellar activity are spots, plages, convective suppression, and other inhomogeneities on the stellar surface. As the star rotates, the active regions move in and out of view, altering the shape of spectral lines and leading to RV variations \citep[e.g.][]{Barnes2011}. These activity signals tend to appear at the stellar rotation  period  and  its  harmonics \citep{Boisse11}. \\
\indent It is therefore crucial to properly account for stellar activity when searching for exoplanets, particularly in systems where the orbital period of a planet is close to the stellar rotation period. The GJ~832 system is reported to host two planets:  an outer jovian planet with a long period orbit of 3660\,days (planet b) \citep{Bailey09}, and an inner planet with a minimum mass of $5.4 \pm 1.0$ $\rm{M}_\oplus$ and an orbital period of $35.68 \pm 0.03$ days (planet c) \citep{Wittenmyer14}. This latter study used 109 RV data points, which we incorporate herein along with our new  RVs.
Subsequent work found a stellar rotation period of $45.7 \pm$ 9.3 days, using the Ca II H\&K lines from the 53 publicly available High Accuracy Radial velocity Planet Searcher (HARPS) spectra \citep{SuarezM15}, also analysed in this work. The  Ca II H\&K derived stellar rotation period and the previously derived GJ~832c period are almost coincidental within the measured uncertainties. Therefore, a rigorous analysis of stellar activity should be performed to determine the origin of this RV signal.   \\
\indent In this work we study both photometric and RV  data of GJ~832 using archival and new observational data.  In Section \ref{sec:properties} we present the  stellar properties, and in Section \ref{sec:obs} we describe the data used in this study. In Section \ref{sec:rotation} we retrieve the stellar rotation, while in Section \ref{sec:rv-analysis} we analyse periodograms and perform Keplerian models. We modelled RV plus activity indicators in Section \ref{sec: rv-act} to consequently update the orbital parameters of the system in Section \ref{sec: 1keplerian-gp}. In Section \ref{sec:discussion} we discuss the possibility of the existence of a planet with an orbital period that is close to the stellar rotation period to finally summarise  our  conclusions in Section \ref{sec:summary-and-concl}.

\section{Stellar properties}
\label{sec:properties}

The main stellar parameters of GJ~832 are listed in Table \ref{tab:gl832}. Studies of the stellar atmosphere \citep[e.g.][]{Fontanela16,Kruczek17,Peacock19,Duvvuri21} have been developed in order to asses the impact of its radiation on the atmosphere of the orbiting planets. These analyses have shown ultraviolet and X-ray fluxes present in  GJ~832, confirming that it is a magnetically active star. Its stellar photosphere, chromosphere, transition region, and corona have been modelled by \citet{Fontanela16}, who find the extreme ultraviolet flux of GJ~832 to be comparable to the active Sun. 

\begin{table*}
\caption{Stellar properties of GJ~832.}
\label{tab:gl832}
\centering
\begin{tabular}{c c c}
\hline\hline
Parameter & Value & Reference \\
\hline 
Spectral Type            & M2V                                                & \cite{SuarezM15} \\
Age                      & 6 $\pm$ 1.5 Gyr                                    & \cite{Guinan16} \\ 
RA (J2000)               &  21h33m33.97s                                      & \cite{Gaia}\\
Dec (J2000)              & -49$^{\circ}$00$^\prime$32.40$^\prime$ $^\prime$   & \cite{Gaia} \\
U [mag]                  & 11.359                                             & \cite{Koen} \\
B [mag]                  & 10.176                                             & \cite{Koen} \\
V [mag]                  & 8.672                                              & \cite{Koen} \\
Parallax (mas)           &  201.4073 $\pm$ 0.0429                             &  \cite{Gaia} \\
Distance (pc)            & 4.9651 $\pm$ 0.0011                                & \cite{Sebastian2021}  \\
Mass (M$_{\odot}$)       & 0.45 $\pm$ 0.05                                    & \cite{Bailey09}\\
Radius (R$_{\odot}$)     &  0.499 $\pm$ 0.017                                 & \cite{Houdebine10} \\
T$_\mathrm{eff}$ (K)     & 3580 $\pm$ 68                                      & \cite{Maldonado15}\\ 
Fe/H (dex)               & -0.16 $\pm$ 0.09                                   & \cite{Maldonado15}\\
log g (cgs)              & 4.82 $\pm$ 0.05                                    & \cite{Maldonado15} \\
log$_{10}$ R'$_{\rm HK}$ & -5.21 $\pm$ 0.07                                   & \cite{SuarezM15} \\
                         &  -5.222 $\pm$ 0.071                                &This work following \cite{Astudillo-Defru2017act} \\
P$_{rot}$ (days) & 37.5 $\substack{+1.4 \\ -1.5}$                          & This work \\
\hline
\end{tabular}
\end{table*}

\section{Observations}
\label{sec:obs}
\subsection{High-resolution spectroscopic data}
\label{sec:rv-data}
This work makes use
of data from HARPS \citep[][]{Mayor03}, the University College London Echelle Spectrograph \citep[UCLES;][]{Diego90}, and the Planet Finding Spectrograph \citep[PFS;][]{Crane06}. HARPS data are available as raw images and reduced spectra, while we accessed UCLES and PFS data only as RV time series. We used a total of 227 RV data points for GJ~832. A summary of each dataset is shown in Table \ref{tab:RV-data}.

\begin{table}
\caption{Main properties of the different RV datasets.}
\label{tab:RV-data}
\centering
\begin{tabular}{c c c c }
\hline\hline
         Properties      & HARPS & UCLES & PFS \\
         \hline
         Years & 2003-2020& 1998-2013 & 2011-2013 \\
         Time-span (days)& 5858 & 5465 & 818 \\
         Mean error (ms$^{-1}$)&0.50 & 2.59 & 0.9 \\
         N$_{\mathrm{data}}$& 172& 39& 16 \\
\hline
\end{tabular}
\end{table}
\subsubsection{HARPS}
HARPS is mounted on the 3.6m telescope at La Silla Observatory located in Chile. The instrument has a spectral resolution of 115,000 and a wavelength coverage between 378 nm and 691 nm. 
We incorporated a total of 172 spectra from the HARPS spectrograph with a time span of 5858 days, from 2003 to 2020.  Out of the entire dataset, 119 entries correspond to new data, and 110 of these were obtained after the fibre upgrade \citep[HARPS+;][]{LoCurto15}. The new data were taken from HARPS runs: 072.C-0488(E), 183.C-0972(A) and  198.C-0836(A) (63 measurements), and 0104.C-0863(A) from the RedDots programme \citep{RedDots2020} (56 measurements).

We computed RVs for the full HARPS dataset using NAIRA \citep[New Algorithm to InferRAdial-velocities;][]{Astudillo-Defru2017b}. This algorithm uses spectra to built a high signal-to-noise ratio (S/N) stellar template and telluric template. The latter is used to mask out tellurics and the former is used to determine the RV of each individual spectrum by maximising the likelihood of the value of the Doppler shift. The template matching approach to compute precise RVs for M dwarfs has shown significant improvements over the cross-correlation function used in the HARPS data reduction software \citep[e.g.][]{AngladaEscude-Butler12,Zechmeister18}. The RV uncertainty was computed following \cite{Bouchy2001}, resulting in an average uncertainty of 0.50 m\,s$^{-1}$ in the range from 0.25 m\,s$^{-1}$ to 2.78 m\,s$^{-1}$. The average S/N at 612~nm (HARPS order 60) corresponds to 87.

We computed spectroscopic activity tracers from HARPS data. The S index, defined from the emission lines of Ca II H \& K \citep{Vaughan}, was computed following the method detailed in \citet{Astudillo-Defru2017act}. For H$\alpha$ we used the method described by \citet{Gomes2011} and for Na D we followed \citet{Astudillo-Defru2017b}. As for H$\beta$ and  H$\gamma$, they  were computed by using the following procedure:
\begin{equation}
    \mathrm{index}= \frac{\mathrm{C}}{\mathrm{R+V}}
,\end{equation}
where C corresponds to the integrated flux in the spectral line and R and V are the two continuum domains. For H$\beta$, the spectral ranges (in nm) for C, R, and V  correspond to [4861.04, 4861.6], [4862.6, 4867.2], and [4855.04, 4860.04], respectively. Whereas for  H$\gamma$ we used C: [4340.162,4340.762],  R:[4342.0,4344.0], and V:[4333.6,4336.8].  The instrument pipeline provides the contrast, full-width at half-maximum, and bisector of the cross-correlation function.
\subsubsection{UCLES}
UCLES is mounted on the Anglo-Australian Telescope located at the Australian Astronomical Observatory. Its spectral resolution is 50,000 covering a wavelength range between 300 nm and 1100 nm. We added a total of 39 data points from UCLES obtained in a time span of 5465\,days (from 1998 to 2013), and they are are publicly available \citep{Wittenmyer14} with a mean error of 2.59 m\,s$^{-1}$.
\subsubsection{PFS}
\label{sec:pfs}
 We also included data from the PFS spectrograph installed in the 6.5\,m Magellan Clay telescope located at Las Campanas Observatory in Chile. This echelle spectropgrah has a  resolution of 80,000 over a wavelength range between 391 nm and 731 nm.  A total of 16 PFS measurements are available \citep{Wittenmyer14} over 818 days (from 2011 to 2013), with an average RV uncertainty of 0.9 ms$^{-1}$. 
\subsection{Photometric data}
\label{sec:phot-data}
\begin{figure}
    \centering
    \includegraphics[width=\hsize]{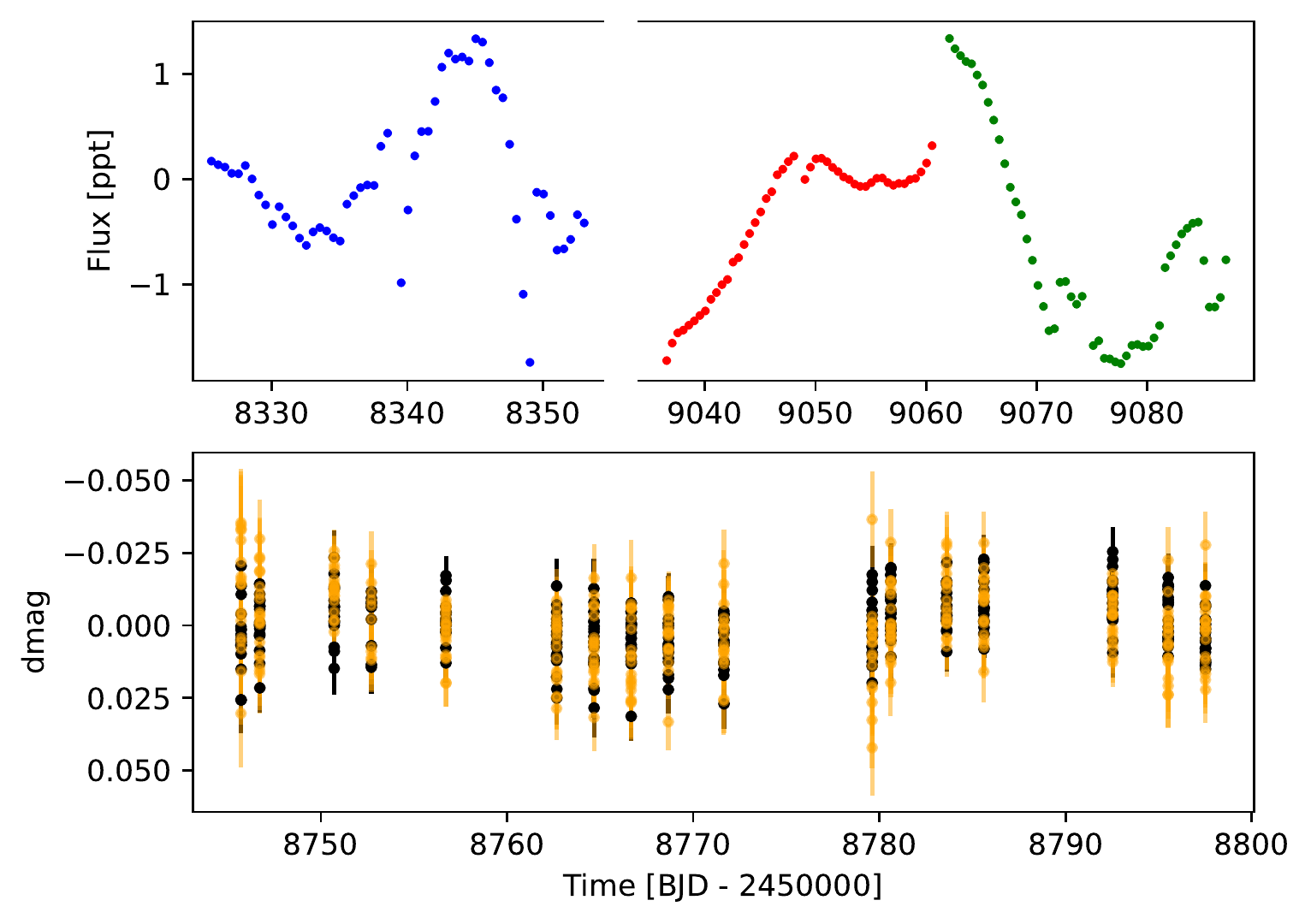}
    \caption{Photometric time series of GJ~832. \textit{Top panel:}  TESS (SAP) data from sectors 1 (blue), 27 (red), and 28 (green). The flux is given in parts per thousand (ppt). \textit{Bottom panel:} ASH2 data using $B$ (black) and $V$ (orange) filters, which are given in differential magnitudes. 
    }
    \label{fig:phot_timeseries}
\end{figure}

We analysed photometric data from TESS \citep{TESS} and ASH2, which are displayed in Figure~\ref{fig:phot_timeseries}. 

\subsubsection{ASH2}
We include photometric CCD observations of GJ~832 collected with the robotic 40-cm telescope ASH2 located at the SPACEOBS observatory in San Pedro de Atacama, Chile. The 40\,cm robotic telescope is  operated by the Instituto de Astrof\'\i fica de Andaluc\'\i a (IAA, CSIC), which  is equipped with a CCD camera STL11000 2.7K$\times$4K, FOV 54$\times$82 arcmin. A total of 633 measurements were collected, of which 316 were taken with the $B$ filter and 317 with the $V$ filter. The data were obtained over 17 nights spanning 52 days during the period from September to November 2019.  The typical exposure times are 15 and 8 seconds for both filters. The typical photometric aperture radius used in the data reduction is 1.23$^{\prime\prime}$/pixel.

Only subframes with 40\% of the total field of view (FOV)  were used for an effective FOV of 21.6$\times$32.8 arcmin.  All CCD measurements were obtained by the method of synthetic aperture photometry without pixel binning. Each CCD frame was corrected in a standard way for dark and flat fielding. Different aperture sizes were also tested in order to choose the one that produces the light curve with the least dispersion. The nearby and relatively bright stars showing the lowest root mean square (RMS) flux scatter were selected as reference stars. 
\subsubsection{TESS}
TESS observed GJ\,382 in three Sectors, 1, 27, and 28, making a time span of 761 days. We made use of the light curves available on the Mikulski Archive for Space Telescopes (MAST \footnote{\url{https://mast.stsci.edu/portal/Mashup/Clients/Mast/Portal.html}}). GJ 832 does not have any transiting planets, but with this photometry data we can study the stellar variability. Since the pre-search data conditioning simple aperture photometry (PDCSAP) light curves are cleaned from long-term variabilities, we used simple aperture photometry (SAP) instead. To compensate for offsets between each SAP light curve, we used the light curves from each sector independently, making a total of three separate light curves. 

To correct for systematics, we made use of the co-trending basis vectors (CBVs) that are generated in the PDC component of the TESS pipeline for each sector and CCD. These can be used to remove the most common systematic trends. The CBV correction is made in {\tt lightkurve} \citep{2018ascl.soft12013L} using the {\tt CBVCorrector} method in combination with the single-scale basis vector type which better preserves long-term signals. 

\section{Rotation period}
\label{sec:rotation}
\subsection{Spectroscopic rotation period}
\label{sec:rot-spec}
\begin{figure}
    \centering
    \includegraphics[width=\hsize]{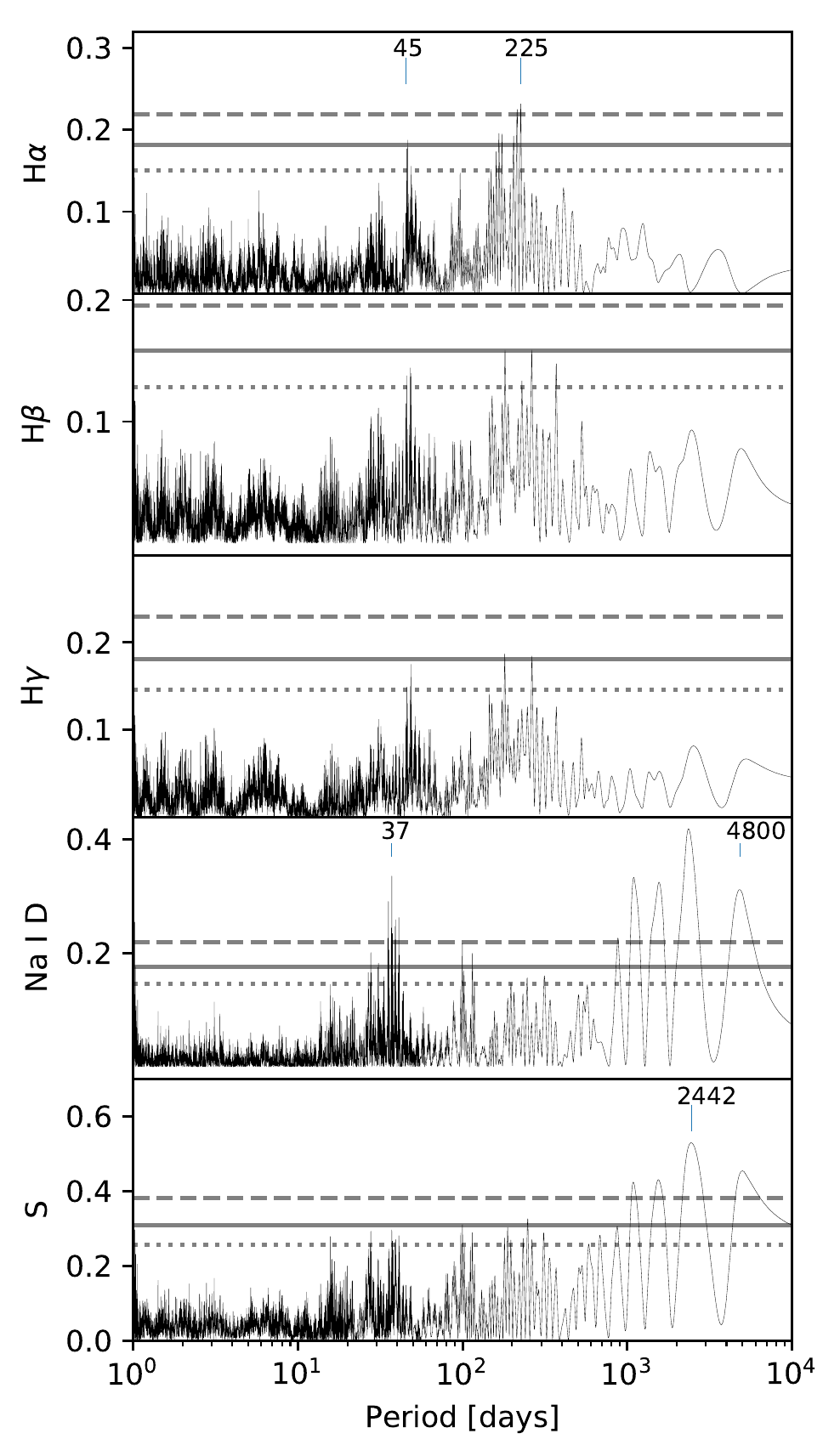}
    \caption{GLS periodograms of the activity tracers.  From top to bottom: H$\alpha$, H$\beta$, H$\gamma$, Na D I, and S index. The dashed, solid, and dotted horizontal lines represent the 0.3\%, 4.6\%, and 31.7\% FAP levels, corresponding to a 3$\sigma$, 2$\sigma$, and 1$\sigma$ detection threshold, respectively.}
    \label{fig:gls-index}
  \end{figure}

We calculated the following activity indicators from HARPS data: H$\alpha$, H$\beta$, H$\gamma$, Na I D, and S index.\ The latter is widely used to study activity \citep[e.g.][]{Duncan91}. Their generalised Lomb-Scargle (GLS) periodogram \citep[][]{Zech19} is displayed in Figure~\ref{fig:gls-index}. For the H$\alpha$, H$\beta$, and H$\gamma$ indices, there are strong peaks at 45d and around 200d. As for the Na I D and S index, the largest peaks correspond to large periods between 1000 and 4000 days, and this may possibly be due to a long-term variation in the activity (stellar cycle type); the study of which is not within the scope of this article. There is also a significant peak at 37d in the sodium index, which is also present in the S index (2$\sigma$).

\begin{figure}
\sidecaption
\includegraphics[width=1.1\hsize]{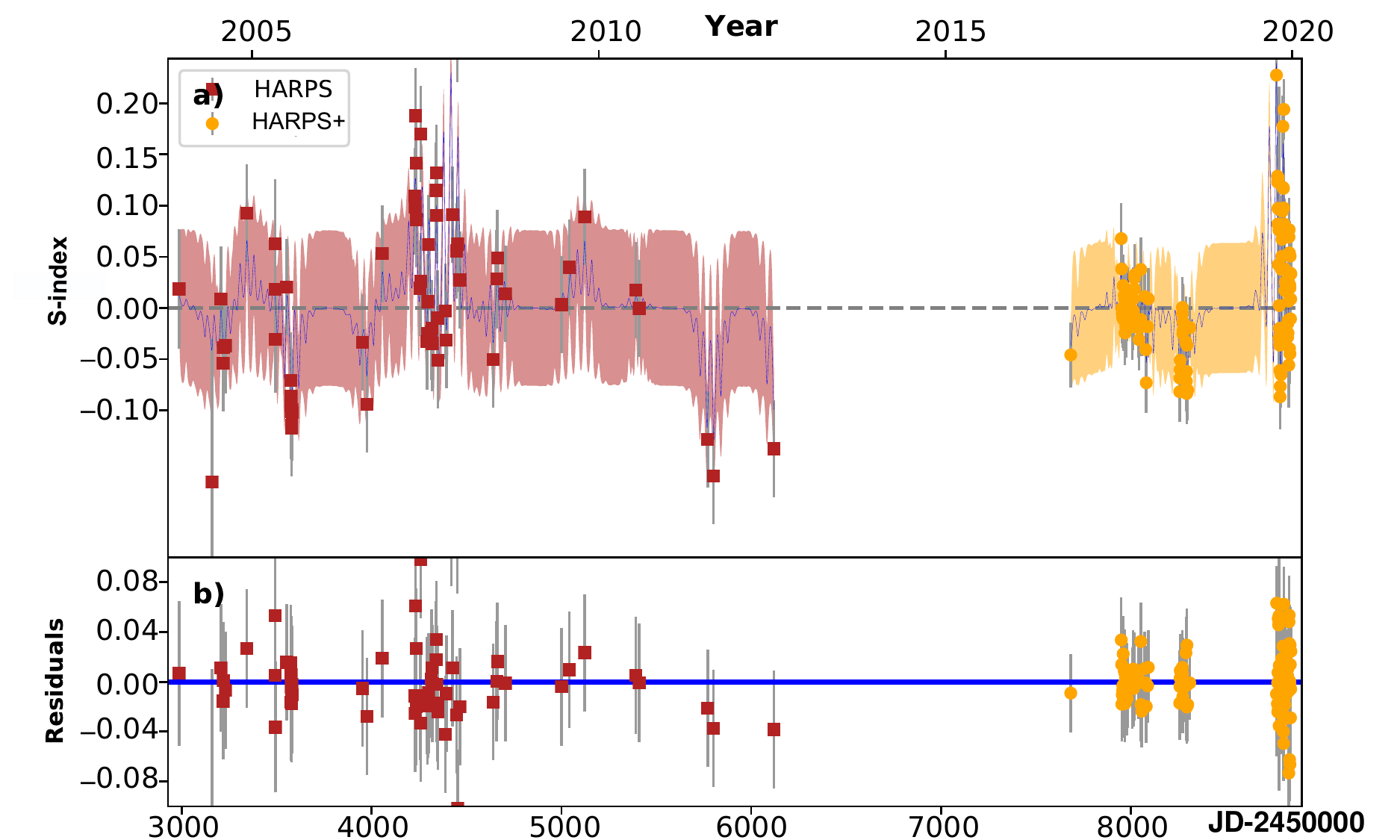}
  \caption{S index time series modelled with a GP using a quasi-periodic kernel.  Both HARPS (red squares) and HARPS+ (yellow circles) data are shown.  \textit{a)} Model showing the GP resulting from the median of the hyper-parameters' posteriors (blue curve). The coloured zone depicts the model 1-$\sigma$ confidence. Error bars account for the white noise included in the fit. \textit{b)} Residual of the fit.}
\label{fig:Sindex-GP}
\end{figure}

The stellar rotation period can be derived from S-index time series. As previous studies have shown \citep[e.g.][]{Haywood14, Grunblatt,Rajpaul}, the S index can be modelled using a Gaussian process (GP) regression. In particular, the quasi-periodic kernel in a GP has a hyper-parameter -- the periodic component -- attributable to the stellar rotation period. This kernel is described in \hyperref[sec:appendix-GP]{Appendix A}. We used the GP modelling capability of  \href{https://radvel.readthedocs.io/en/latest/}{\textsc{RadVel}} \citep{Fulton18}  to perform this modelling. 
The priors on the hyperparameters correspond to the following uniform priors: in a range of 0 and 0.3 for the amplitude of correlations ($\eta_1$); from 0 to 4000 days for the aperiodic timescale decay ($\eta_2$); between 1 and 200 days for the periodic component ($\eta_3$); and from 0 to 7 for the periodic timescale ($\eta_4$). 
Readers can refer to \hyperref[sec:appendix-rot]{Appendix B} and  \hyperref[sec:MCMC]{Appendix D} for more details on the priors and the  Markov chain Monte Carlo (MCMC) results, respectively. Figure~\ref{fig:Sindex-GP} shows the time series of the S index with the GP model. The periodic component of the quasi-periodic kernel has a value of 37.5  $\substack{+1.4 \\ -1.5}$ days (see Fig. \ref{fig:radvel-corner}), which is extremely close to the orbital period of putative planet c. Since stellar activity varies over time (and therefore also the activity indicators), we also modelled each HARPS campaign separately and retrieved a clear detection around 37 days for each dataset: 37.13 $\pm$ 2.03 days for  HARPS and 35.99 $\pm$ 1.66 days for HARPS+, which is consistent with the periodicity obtained using the entire time span. From the S-index fit, the measurement of the stellar rotation period is improved by a factor of 6 with respect of the reported value by \citet[][]{SuarezM15}.

We also modelled the other activity tracers (H$\alpha$, H$\beta$, H$\gamma$, and Na D I) with a GP, using the same priors as in the S-index model. We found a clear rotation detection on the Na DI index, giving a result of 36.6 $\pm$ 1.2 days, which is in agreement with the S-index measurement. For the remaining activity indices, we could not retrieve a clear detection of the rotation period (see Fig.~\ref{fig:gls-index}).

Since the time span is larger for the RV data than for the photometry and because the S index is widely used for modelling stellar activity, we used this measurement for the final value of the stellar rotation. It is confirmed with the photometry data, as is shown in the following sub-section.
\subsection{Photometric confirmation}
\begin{figure}
    \centering
    \includegraphics[width=\hsize]{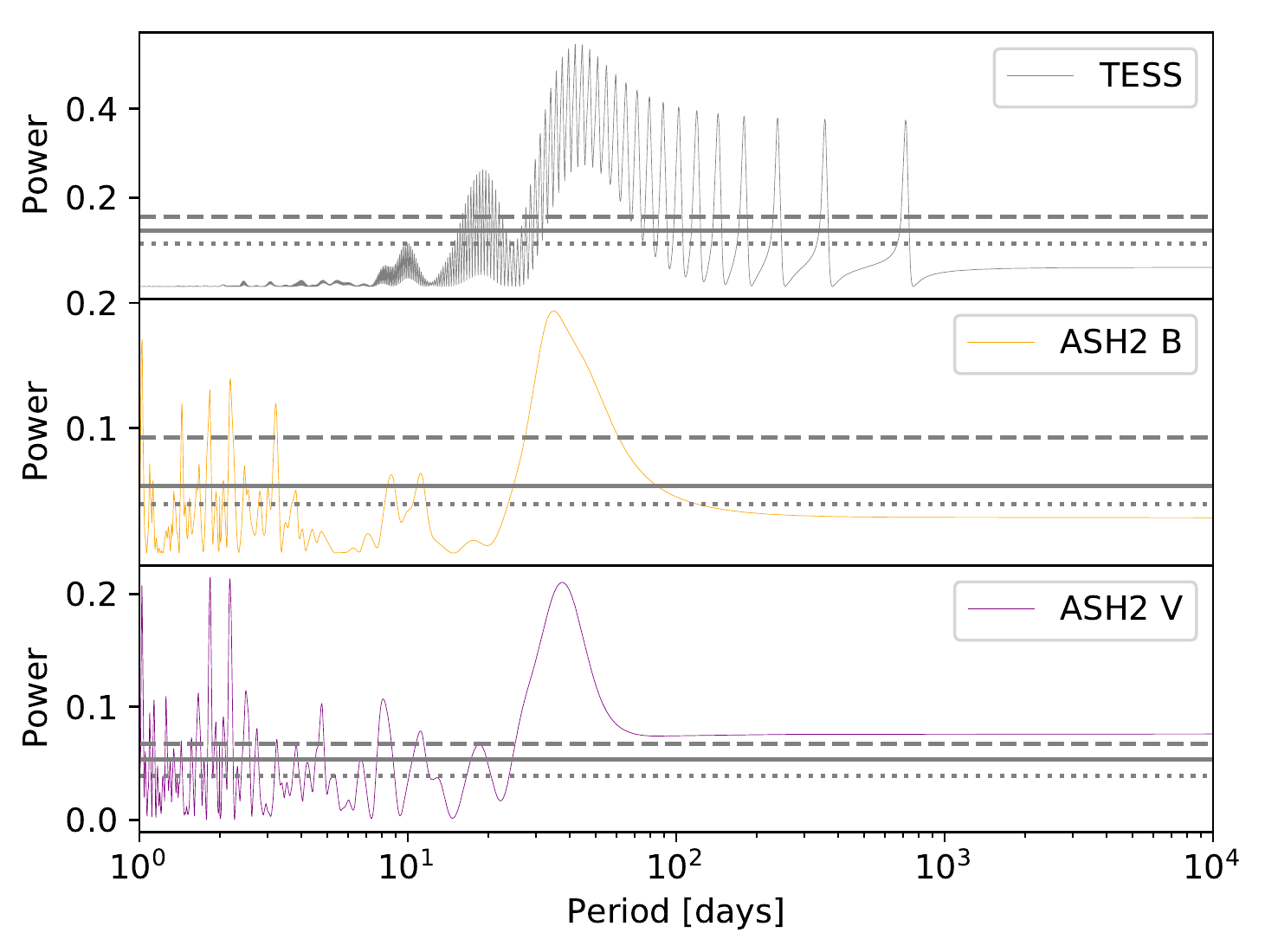}
    \caption{GLS periodograms of GJ~832 from TESS (top) and ASH2 data using the $B$ (center) and $V$ (bottom) filters. The dashed, solid, and dotted horizontal lines are the same as in Fig. \ref{fig:gls-index}.}
    \label{fig:gls_phot}
\end{figure}

 We computed the GLS periodogram of the photometric data (TESS and ASH2), as shown in Figure~\ref{fig:gls_phot}. There is a dominant peak around 35 days for every dataset, which could be the stellar rotation period as expected from the spectroscopic rotation period measurement. 
 
\begin{figure}
    \centering
    \includegraphics[width=\hsize]{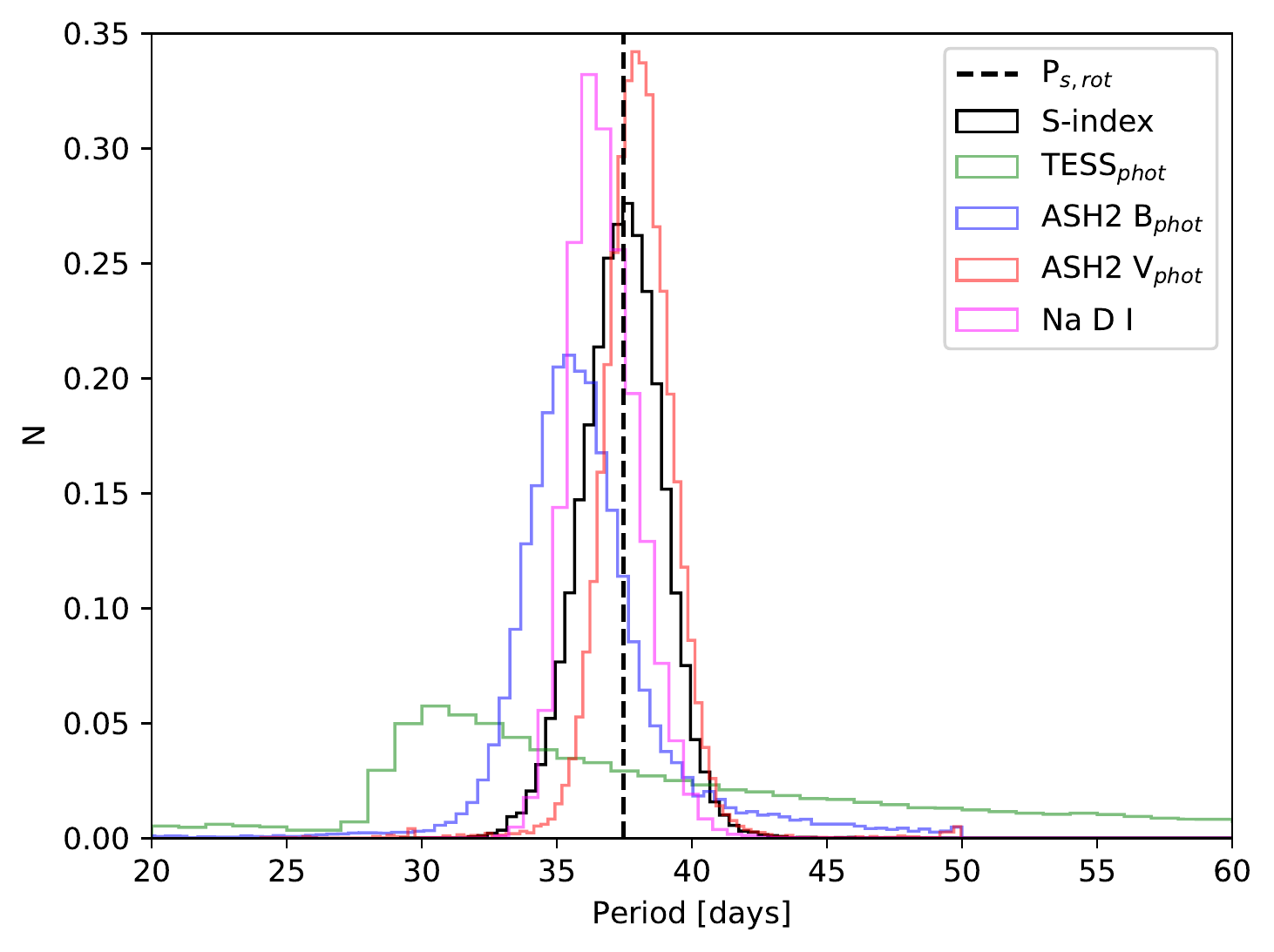}
    \caption{Histogram of the periodic component of the quasi-periodic kernel $\eta_3$, equivalent to the stellar rotation period. The vertical dashed line represents our measured value from the S index. We also depict distributions of the Na D I index and the photometric data from TESS and ASH2. }
    \label{fig:n3_histogram}
\end{figure}

 To corroborate the rotation period, we also performed a GP regression on each photometric dataset. As with the activity analysis, we also made use of the quasi-periodic kernel.  Further details as well as the posterior distribution of the parameters are given in \hyperref[sec:appendix-GP]{Appendix A} and \hyperref[sec:appendix-rot]{Appendix B}. Their resulting period distributions are shown in Figure~\ref{fig:n3_histogram} along with the Na D I and S-index histograms.  The periodic component of the ASH2 data is consistent with the stellar rotation period derived from the S index, whereas the TESS data show a wide distribution with a detection at 39.6  $\substack{+29 \\ -8.9}$ days, which is also in agreement with the rotation measurement. A single TESS sector is shorter than the rotation period, while the offset between sectors compromises the ability of the two consecutive sectors to discern the period (Fig. 1). Therefore, the photometric data confirm the result obtained from the S-index analysis. 
 
\section{RV analysis}
\label{sec:rv-analysis}
The RV dataset shows a clear and significant variation reaching 20 \,m\,s$^{-1}$ with a mean error of 0.89\,m\,s$^{-1}$. We computed the GLS periodogram of the RVs, where we observe the most significant peak around 4000 days (planet b) far above the 3$\sigma$ detection threshold (top panel of Figure~\ref{fig:gls}). We modelled the RVs using \href{https://github.com/oscaribv/pyaneti}{\textsc{Pyaneti}}  \citep{Barragan22}. 
\subsection{1-Keplerian model}
\label{sec:1K}
\begin{figure}
    \centering
    \includegraphics[width=\hsize]{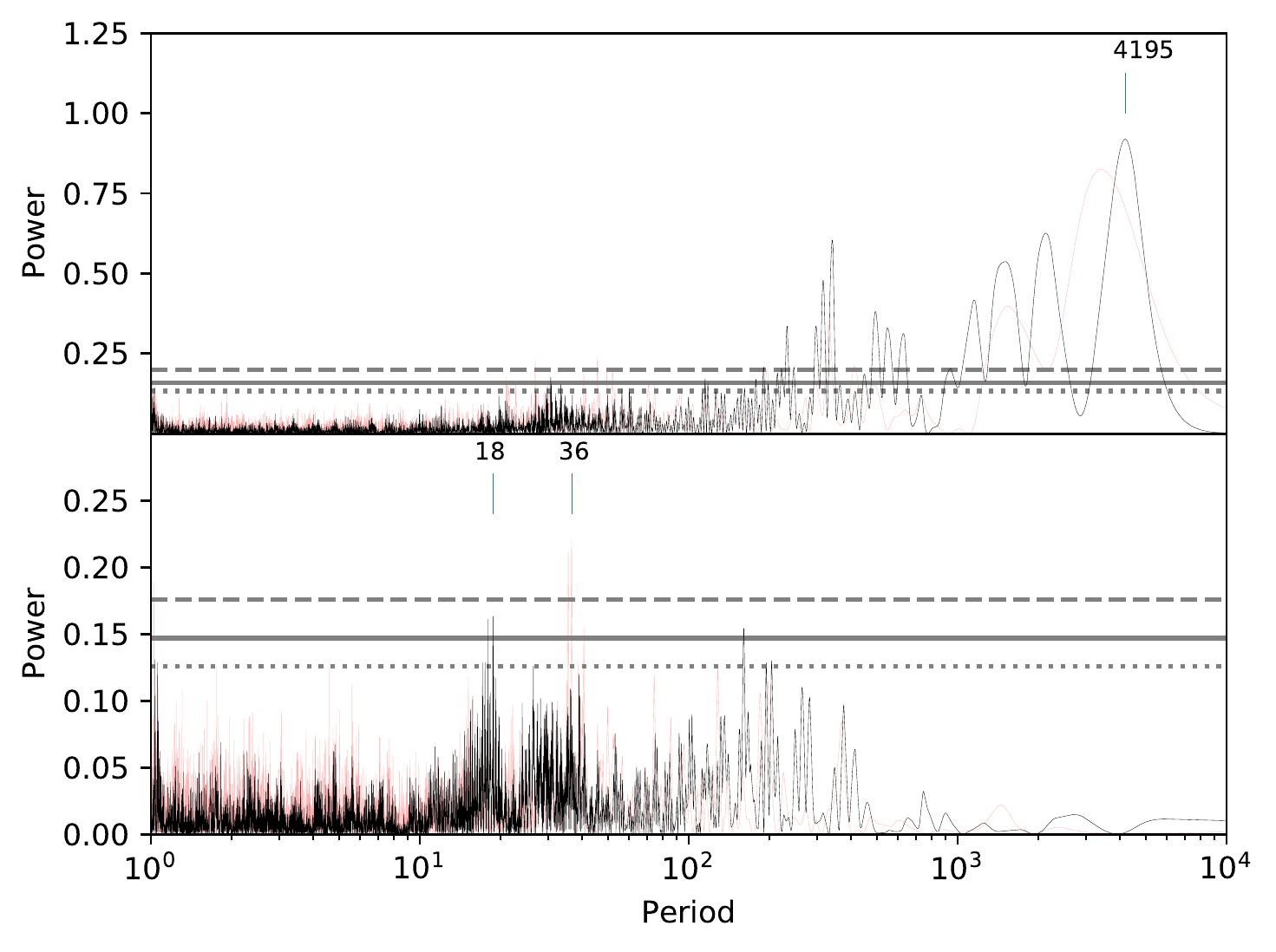}
    \caption{GLS periodograms of GJ 832 from the RVs' time series (top) and the RVs' residual after removing the most significant signal (bottom). The old data are illustrated in red, whereas the new data (which includes 119 new measurements) are shown in black. The dashed, solid, and dotted horizontal lines are the same as in  Fig. \ref{fig:gls-index}.} 
    \label{fig:gls}
\end{figure}

For this model, we used uniform priors for the period around 3500 and 4200 days (see Table~\ref{tab:one-kep} for more details about the priors). The largest peak in the residual periodogram corresponds to 18.7\,days, which is equivalent to half of the stellar rotation found in Section \ref{sec:rot-spec}, whereas the 35d signal is not significant as it lies below 1$\sigma$ (bottom panel of Figure~\ref{fig:gls}). We note that without our new HARPS data, the 35d signal is significant.
 
We analysed the temporal stability of the signals present in RV residual after subtracting the model for planet b. We used the stacked Bayesian General Lomb-Scargle (sBGLS) periodogram formulated by \citet{Mortier2017}. This enables us to discriminate stellar activity signals from planetary signals as active zones may not be stable over time in amplitude and phase because they appear or disappear at different regions of the stellar surface, while planetary signals are stable over time, thus being phase coherent. Hence, the power (or probability) with a planetary origin should always increase as more observations are added into the dataset. 

\begin{figure}
    \centering
    \includegraphics[width=\hsize]{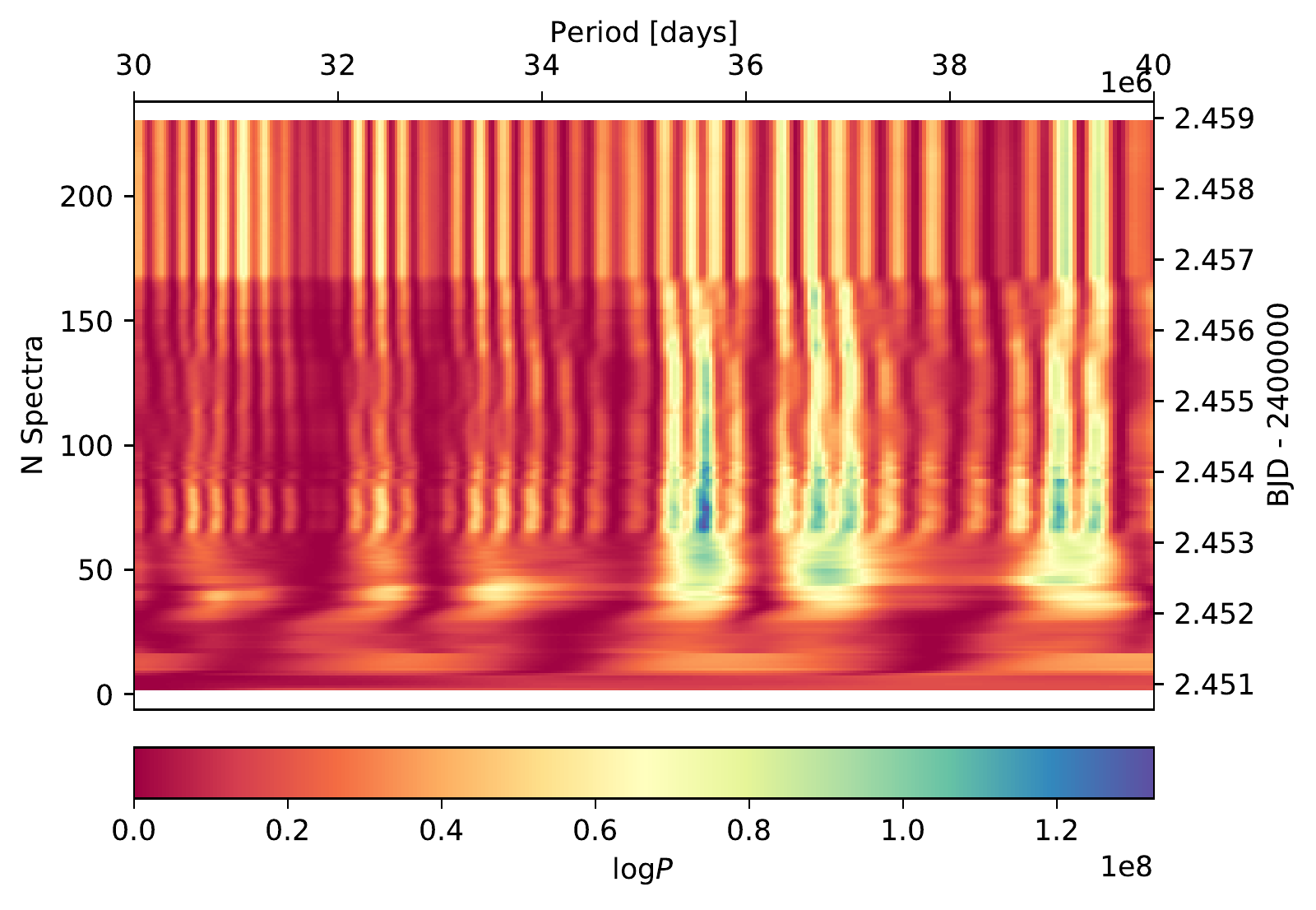}
    \includegraphics[width=\hsize]{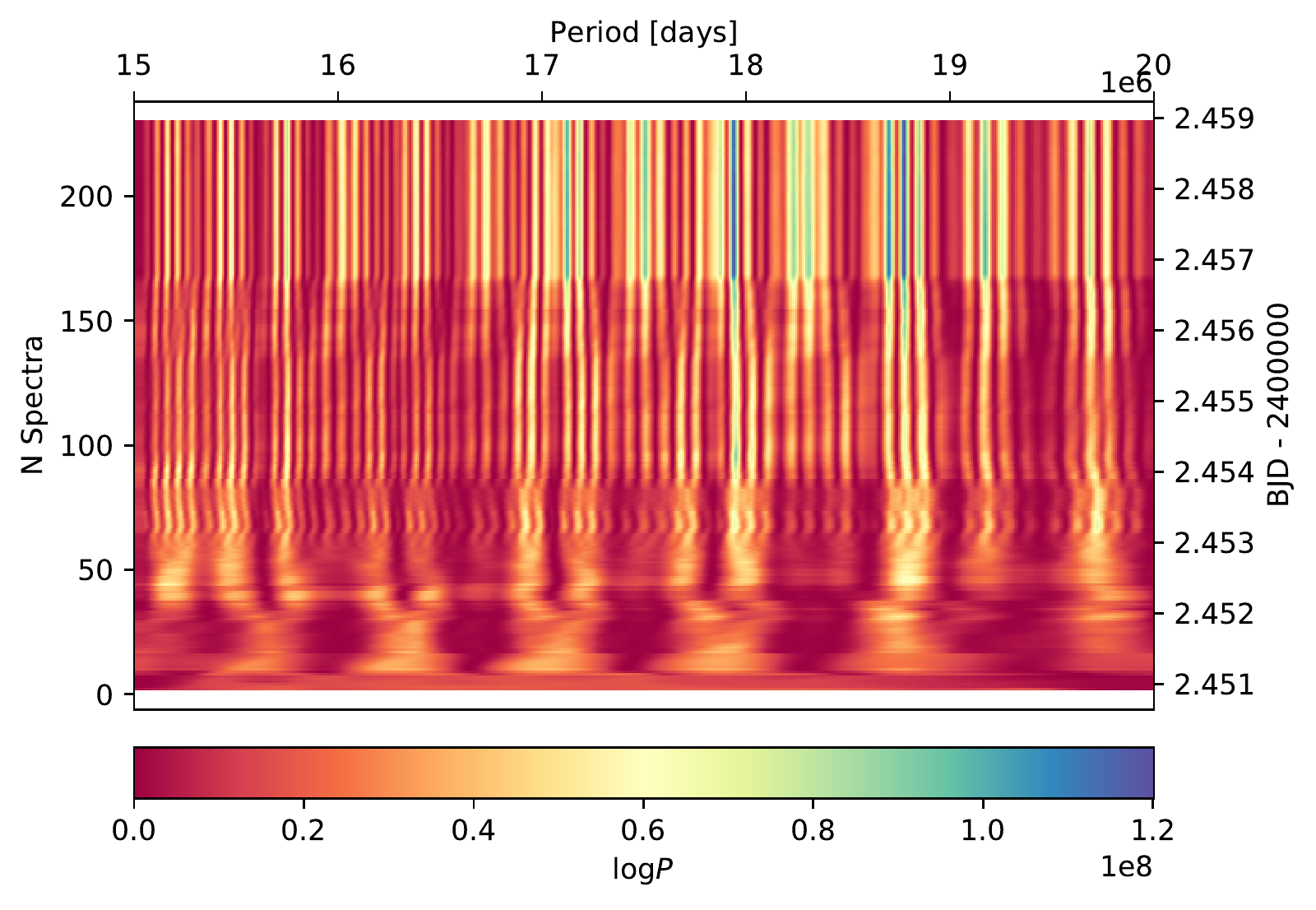}
    \caption{sBGLS periodograms of the RVs' residual (after subtracting the signal of planet b) centred around the 35d signal (top) and around 18 days (bottom), which corresponds to the largest peak on the residual periodogram. The number of observations is plotted against the period, whereas the colour bar indicates the logarithm of the probability. }
    \label{fig:sbgls-res}
\end{figure}

The top panel of Figure~\ref{fig:sbgls-res} shows the sBGLS periodogram of RV residual around  35 days. We note that as the number of observations grows, the probability around 35\,d does not steadily increase; it starts to lose its significance when the last data points are added. The bottom panel shows the sBGLS periodogram around 18 days, the largest peak in the residual periodogram. Here we see the contrary: at the end of the dataset, this signal becomes more stable, which is in accordance with the GLS periodogram of the residual.  The coherence of this latter signal favours a second planet at 18\,d rather than at 35\,d. A further analysis is performed in a 2-Keplerian model in the following section.

\subsection{2-Keplerian model}
\label{sec:2-kep}
 We also performed a 2-Keplerian model fit to check if the 35d and 18d signal could have a planetary origin. In this model, we considered the planet in the wide orbit (planet b)  as one Keplerian, using the same priors as the 1-Keplerian model from Section \ref{sec:1K},  and then we added a second Keplerian using uniform priors on the period from 2 to 50 days. The rest of the priors are listed in Table \ref{tab:2kep}.
 
The best orbital solutions for this model gives an unclear detection (broad distribution and multiple peaks) for the second Keplerian, with a periodicity of $28.09^{+7.54}_{-9.36}$ days. Therefore, a second Keplerian with a period of 18 or 15 days is not preferred for this model, suggesting that these signals do not correspond to planets. Motivated by this, simultaneous RV and activity indicator analyses are performed in the following section to determine whether these signals are related to stellar activity. 

\section{RV and activity indicators}
\label{sec: rv-act}
We analysed the RVs alongside the activity indices H$\alpha$, Na D I, and the S index; each were analysed independently.  Since we only have the activity indices from HARPS, we only used the RV measurements from this instrument  to perform a simultaneous modelling with the same time series. The number of data points and the high precision of these measurements make this approach informative.  We used 344 HARPS data points (172 RV measurements and the corresponding 172 measurements of activity data).

We explored the full (hyper-)parameter space with the publicly available Monte Carlo (MC) nested sampler and Bayesian inference tool \textsc{MultiNest v3.10}  \citep[e.g.][]{Feroz2019}, through the \textsc{pyMultiNest} wrapper \citep{Buchner2014}. We set up 1000 live points, a sampling efficiency of 0.3, and a tolerance on the Bayesian evidence of 0.5. To perform the GP regression, we used the package \textsc{george} \citep{ambikasaran14}, using the QP kernel. Model comparison was performed using the logarithm of the Bayesian evidence $\ln\mathcal{Z}$ provided by \textsc{MultiNest}. The model comparison was done between the models with the same dataset but with different numbers of Keplerians. In this sense, we fitted  one Keplerian with activity indices and then performed another model adding a second Keplerian (using the same priors).

To reduce the computation time, we used two (out of four) hyper-parameters in common: the rotation period (unique value for the star) and the evolutionary time scale. The latter could be different for RVs and activity indices, but it does not impact the modelling of the planetary signal. The priors of these models are listed in Table \ref{tab:kep-act}. Table \ref{tab:comparison} compares the results of this fitting.  We see that for the models using the Na D I and H$\alpha$ indices, there is strong evidence against the second Keplerian ($\Delta\ln\mathcal{Z}$= 5.4). As for the model containing the S index, it still indicates that the 1-Keplerian model has to be strongly preferred ($\Delta\ln\mathcal{Z}$= 4.8).

\begin{table*}
\caption{Model comparison table from the simultaneous RV and activity index fits.}
\label{tab:comparison}
\centering
\begin{tabular}{c| c c c c c }
\hline\hline
Index & Model &  $\ln\mathcal{Z}$ & $\Delta\ln\mathcal{Z}$& $P_{\mathbf{rot}}$ (days) & P$_2$ (days)\\
\hline
       & 1-Keplerian & 1988.8 &     &$36.86^{+0.84}_{-0.81}$ &      \\
Na D I &             &        & 5.4 &                        &       \\
       & 2-Keplerian & 1983.4 &     & $36.91^{+0.79}_{-0.78}$ & $29.28^{+12.76}_{-15.82}$  \\
\hline
       & 1-Keplerian & 1074.1 &     & $36.95^{+0.82}_{-0.83}$  & \\
S index &             &        & 4.8 &                         &\\
       & 2-Keplerian & 1069.3 &      & $36.99^{+0.77}_{-0.80}$ & $29.37^{+12.84}_{-15.25}$  \\
       \hline
       & 1-Keplerian & 1810.3 &        & $35.22^{+7.50}_{-11.95}$ & \\
H$\alpha$ &             &        & 5.4&                           &  \\
       & 2-Keplerian & 1804.9 &       &$35.49^{+7.30}_{-11.32}$  & $29.38^{+12.41}_{-16.20}$ \\
       \hline
\end{tabular}
\end{table*}

Moreover, for all the models, the stellar rotation period is detected around 36 days, as seen in  Table 3. Additionally, in the case of two Keplerians, the models do not succeed at adjusting data where one of the planets has an orbital period periodicity around 35 days. Instead, a broad distribution (with many peaks) around 29 days is detected (P$_2$ in Table 3), which is in agreement with our finding in Section \ref{sec:2-kep} when we modelled two Keplerians (with no GP) for the entire dataset. In all the model cases, the fit of planet b stays unaffected, indicating this is a robust detection.

\section{Best model}
\label{sec: 1keplerian-gp}
\begin{figure*}
    \sidecaption
        \includegraphics[width=1.0\textwidth]{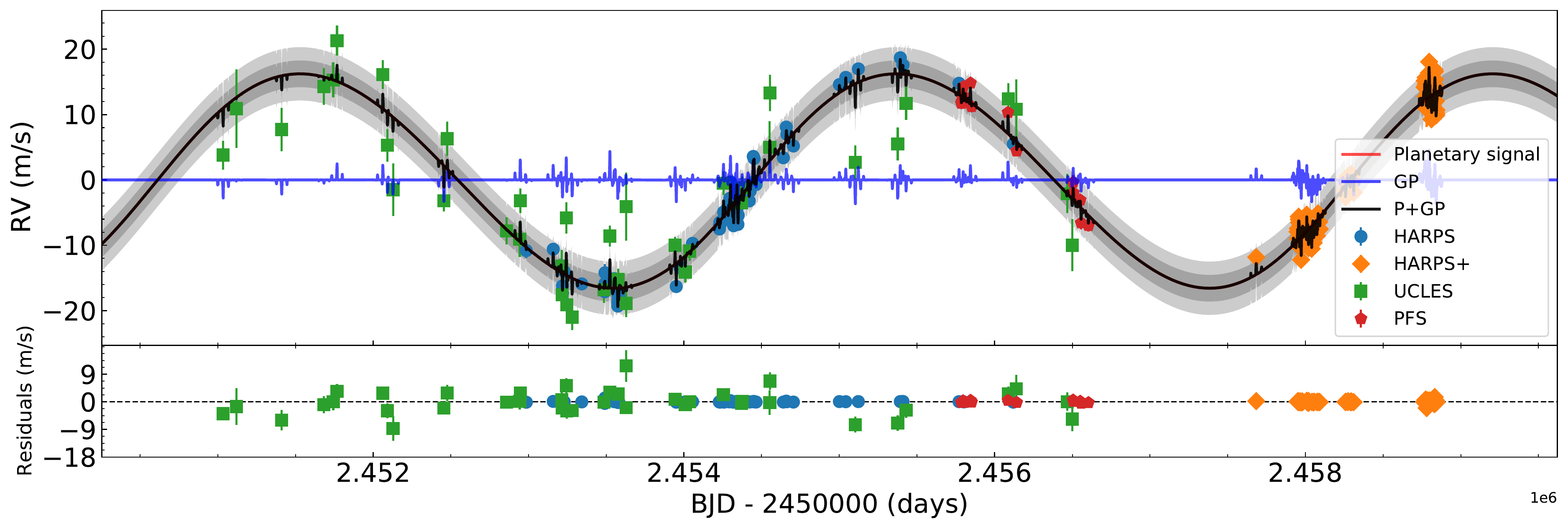}
      \caption{1-Keplerian with GP orbital model for GJ~832 using HARPS (blue), HARPS+ (orange), UCLES (green), and PFS (red) datasets.   \textit{Top panel:} Best fit for the planetary signal and GP (black solid line). The dark grey shaded regions correspond to the 1-and 2-$\sigma$ credible intervals, respectively. \textit{Bottom panel:} Residual of the model.}
    \label{fig:1k-gp-model}
\end{figure*}

Since the model comparison prefers only one Keplerian, we can update the orbital parameters of planet b. We did this by modelling one Keplerian with a GP with \textsc{Pyaneti} \citep{Barragan22}, as we know from the previous section that the GP fits the stellar activity.  We used the entire datasets (not only HARPS) since the UCLES and PFS data increase the time span and therefore the parameters for the wide-orbit planet are better determined.  The model is displayed in Fig. \ref{fig:1k-gp-model} and the orbital values are shown in Table \ref{tab:updated-params}. The priors are listed in Table \ref{tab:1-kep-gp} and the posterior distributions are shown in Fig. \ref{fig:corner-1k-gp}. 

\begin{table}
\caption{Updated orbital solutions of GJ~832 b.}
\label{tab:updated-params}
\centering
\begin{tabular}{c c}
\hline\hline
Parameter                              & Value \\
\hline
Orbital Period $P_b$ (days)            &  $3838.03^{+47.30}_{-49.23}$ \\
Eccentricity $e$                       &   $0.04 \pm 0.02$ \\
Velocity semiamplitude $K$ (ms$^{-1}$) &  $16.41^{+0.35}_{-0.34}$ \\
$\omega$ (radians)                     & 4.46$^{+0.57}_{-0.55}$ \\
Time of conjunction (JD)              & 2456388.70 $^{+48.56}_{-46.34}$ \\
Time of periastron (JD)              & 2457857.81$^{+310.20}_{-3216.97}$ \\
$m$ $\sin{i}$ (M$_{\rm Jup}$)          & 0.74 $\pm$ 0.06\\
\hline
\end{tabular}
\end{table}

\section{Discussion}
\label{sec:discussion}

Our results differ from previous studies of this system as we added 119 new HARPS spectra, making a total of 172 RV measurements. We also incorporated photometric observations.  \citet{Wittenmyer14} performed their analysis with a total of 109 RV spectra when reporting on planet c. When analysing this set of data, after performing a 1-Keplerian fit to the planet in the outer orbit, we also found a significant (3$\sigma$) signal around 35 days in the residual. However, as we added the new HARPS data, this signal has lost its significance and does not even reach the detection threshold (bottom panel of Fig. \ref{fig:gls}) and it is only with the new HARPS RV data that we are able to determine the true origin of the 35d signal. This is particularly noticeable for the addition of the RedDots data (56 RV measurements) which were taken with approximately nightly cadence to minimise sources of correlated noise and to be able to quantify the evolution of stellar activity features \citep[for more details, see][]{Jeffers2020}.

Before we come to a final conclusion, we discuss the possibility that we have a planet with an orbital period close to the stellar rotation period. For example, K2-18\,b \citep{K2-18} is an 8.5\,M$_\oplus$ planet in a 32\,d orbit around an M dwarf with a rotational period of 39\,d. With RV measurements alone, the planetary signal is difficult to disentangle from the rotational modulation. Without the transit detection, this planet would have been undetected. A commensurability between planetary orbit and stellar rotation seems to be possible. Similarly, the K2-3 d planet was discovered by transits \citep{Crossfield15} with an orbital period around 44.5d, whereas the stellar rotation period is around 40d \citep{Damasso18}. This planet is also not detectable using RV data \citep[e.g.][]{Almenara15,Damasso18}, being challenging to disentangle between the signals from stellar activity and the planet, which is a different scenario than GJ~832 c. 

Another case corresponds to HD\,192263 b \citep{Santos2000},  a giant planet discovered by RV with an orbital period around 24d. This planet was discarded due to the proximity of the planetary orbital period and the stellar rotation \citep{Henry2002}, but it was re-analysed with photometry and bisector measurements \citep{Santos2003}, corroborating the presence of the planet. The RV variation showed stability (which is not the case for GJ~832 c) and the photometry showed variability. The planetary orbital period and stellar variability was disentangled \citep{Dragomir12}, attributing a photometric variability of 23d to the stellar rotation. 

A recent example is AU\,Mic\,b with a 7:4 commensurability \citep{AU_Mic}. \citet{WalkowiczBasri} also found an over density of 1:1 and 2:1 ratios between rotational and orbital periods in the Kepler sample, but only for planets with a radius above about 6\,R$_\oplus$. The radius of the planet candidate GJ\,832\,c is probably below that threshold, which makes it unlikely that we have a situation comparable to this case.

Furthermore, both \citet{Vanderburg16} and \citet{Newton16} show that the RV jitter from stellar rotation coincides with the periodrange of Keplerian orbits in the HZ around M dwarfs. This  is indeed the case for the putative planet GJ~832 c. 

\section{Summary and conclusions}
\label{sec:summary-and-concl}
We have collected a significantly expanded dataset for GJ~832, comprised of new RV, activity, and photometric data. We have thus used several statistical tools to study the exoplanetary system, its rotation, and stellar activity. We summarise
our results below.

\begin{itemize}
    \item We measured the stellar rotation period from the spectroscopic activity indices, with a resulting value of 37.5 $^{+1.4}_{-1.5}$ days. This overlaps  within uncertainties with the 35\,d signal reported to be a planet. This result agrees with the photometric data. 
    \item The GLS periodogram from the RV residual detects a significant signal at 18.7 days, which corresponds to half of the rotation period.
    \item The residual stacked Bayesian GLS periodogram shows the 35d signal is incoherent and does not persist, contrary to the expected behaviour of a signal induced by a planetary companion.
    \item The 2-Keplerian models do not find a planet with a periodicity around 35 days.
    \item The RV and activity index models (using HARPS data) prefers a 1-Keplerian model under a Bayesian framework.
\end{itemize}
These results provide a new characterisation and interpretation of stellar activity features for GJ 832. They allow us to conclude with confidence that the previously reported planet corresponding to the
35d signal is an artefact of stellar activity.

\begin{acknowledgements}
We thank the referee for the constructive comments that improved the quality of the manuscript.
P.G acknowledges research funding from CONICYT project 22181925 and from the Deutsche Forschungsgemeinschaft (DR\,281/39-1).  N. A.-D. acknowledges the support of FONDECYT project 3180063. SVJ acknowledges the support of the DFG priority programme SPP 1992 “Exploring the Diversity of Extrasolar Planets (JE 701/5-1). 
F.D.S acknowledges support from
a Marie Curie Action of the European Union (grant agreement 101030103).
R.E.M. gratefully acknowledges support by the ANID BASAL projects ACE210002 and FB210003 and FONDECYT 1190621. YT acknowledges the support of DFG priority program SPP 1992 “Exploring the Diversity of Extrasolar Planets” (TS 356/3-1). J.R. Barnes and C.A. Haswell are funded by STFC under consolidated grant ST/T000295/1.

We acknowledge financial support from the Spanish Agencia Estatal de Investigaci\'on of the Ministerio de Ciencia, Innovación y Universidades through projects PID2019-109522GB-C52, PID2019-107061GB-C64,
PID2019-110689RB-100 and the Centre of Excellence 'Severo Ochoa' Instituto de Astrof\'\i sica de Andaluc\'\i a (SEV-2017-0709).

We acknowledge the support by FCT - Funda\c{c}\~ao para a Ci\^encia e a Tecnologia through national funds and by FEDER through COMPETE2020 - Programa Operacional Competitividade e Internacionaliza\c{c}\~ao by these grants: UID/FIS/04434/2019; UIDB/04434/2020; UIDP/04434/2020; PTDC/FIS-AST/32113/2017 \& POCI-01-0145-FEDER-032113; PTDC/FISAST /28953/2017 \& POCI-01-0145-FEDER-028953. \\

This work made use of \href{https://radvel.readthedocs.io/en/latest/}{RadVel}
\citep{Fulton18}, \href{https://github.com/oscaribv/pyaneti}{Pyaneti} \citep{Barragan22}, \textsc{MultiNest v3.10} \citep[e.g.][]{Feroz2019}), \textsc{pyMultiNest} wrapper \citep{Buchner2014}, \textsc{george} \citep{ambikasaran14}, and {\tt lightkurve} \citep{2018ascl.soft12013L}.  This research includes  publicly available data from the Mikulski Archive for Space Telescopes (MAST) from the TESS mission, as well as data public data from HARPS, UCLES and PFS.
\end{acknowledgements}


\begin{appendix}
\section{GP model}
\label{sec:appendix-GP}
GPs are used to model stochastic processes with some known properties but unknown functional forms. They allow us to model relationships that are not necessarily linear, and they are suitable to model physical processes since they can cover a wide range of functions that can fit the phenomenon. This is why GPs are appropriate to characterise signals of stellar activity: although there are many unknown parameters, we know that they are (quasi-) periodic as they are modulated by stellar rotation. A quasi-period kernel is used with a covariance matrix given by \\

\begin{equation}
\sum_{ij} = \eta^2_1 \exp \Bigg{[} - \frac{|t_i - t_j|^2}{\eta^2_2} - \frac{ \sin^2(\frac{\pi |t_i - t_j|}{\eta^2_3})}{2\eta^2_4}  \Bigg{]} \thinspace ,\end{equation}
where $\Sigma_{ij}$ represents the element of covariance matrix, that is to say the covariance between observations at $t_i$ and $t_j$. The parameter $\eta_3$ corresponds to the periodic component, which in this case is the stellar rotation. The other parameters are related to the correlation between the data points. The aperiodic timescale decay of the correlations is represented by $\eta_2$, which parameterises the evolutionary timescale of the active regions responsible for the observed periodic modulation. The parameter $\eta_4$ is the periodic scale and $\eta_1$ is the amplitude of the correlations. 
\section{Further information on the rotation period derivation}
\label{sec:appendix-rot}

\subsection{S index}

\begin{table}[tbh]
\caption{Priors used in the S-index model with a GP.}
\label{tab:s-index-gp}
\centering
\begin{tabular}{c c c }
\hline\hline
Parameter& Prior& Range \\
\hline
$\sigma_{\rm harps}$  & Uniform & [0.0, 0.3]\\ 
$\sigma_{\rm harps+}$ & Uniform & [0.0, 0.3]\\
$\eta_{1}$            & Uniform & [0.0, 0.3]\\ 
$\eta_{2}$            & Uniform & [0.0, 4000.0]\\
$\eta_{3}$            & Uniform & [1.0, 200.0]\\
$\eta_{4}$            & Uniform & [0.0, 7.0]\\
\hline
\end{tabular}
\end{table}
\subsection{Photometry}
\begin{table}[tbh]
\caption{Priors used in the ASH2 GP model. The data were modelled separately (due to the different filters), but we used the same priors for both ASH2 models. }
\label{tab:ash2-bounds}
\centering
\begin{tabular}{c c c }
\hline\hline
Parameter& Prior& Range \\
\hline
$\eta_{1}$ & Uniform & [0.0, 5.0]\\ 
$\eta_{2}$            & Uniform & [0.0, 2000.0]\\
$\eta_{3}$            & Uniform & [10.0, 100.0]\\
$\eta_{4}$            & Uniform & [0.0, 2.0]\\
\hline
\end{tabular}
\end{table}

\begin{table}[tbh]
\caption{Priors used in the TESS GP model.}
\label{tab:tess-bounds}
\centering
\begin{tabular}{c c c }
\hline\hline
Parameter& Prior& Range \\
\hline
$\sigma_{\rm tess01}$ & Uniform & [0.0, 0.5]\\ 
$\sigma_{\rm tess27}$ & Uniform & [0.0, 0.5]\\
$\sigma_{\rm tess28}$ & Uniform & [0.0, 0.5]\\
$\eta_{1}$            & Uniform & [0.0, 5.0]\\ 
$\eta_{2}$            & Uniform & 40.0\\
$\eta_{3}$            & Uniform & 120\\
$\eta_{4}$            & Uniform & 2.00\\
\hline
\end{tabular}
\end{table}

\newpage

\section{Details on the Keplerian models}
\label{sec:appendix-one-kep}

\
\begin{table}
\caption{Priors of the 1-Keplerian model.}
\label{tab:one-kep}
\centering
\begin{tabular}{c c c}
Parameter& Prior& Range \\
\hline\hline
$P_{b}$        & Uniform & [3500, 4200]  \\ 
$T\rm{conj}_{b}$ & Uniform & [2456320, 2456500]   \\
$e_{b}$         & Uniform & [0.0, 0.1] \\ 
$\omega_{b}$      & Uniform  & [0, 2$\pi$] \\
\\
$K $    [km/s]    & Uniform &  [0.0,1.0] \\ 
\hline
\end{tabular}
\end{table}

\begin{table}[tbh]
\caption{Priors of the 2-Keplerian model.}
\label{tab:2kep}
\centering
\begin{tabular}{c c c }
\hline\hline
Parameter& Prior& Range \\
\hline
$P_{b}$        & Uniform  & [3500, 4200] \\
$T\rm{conj}_{b}$     & Uniform  & [2456320, 2456500] \\
$e_{b}$               & Uniform  & [0.0, 0.1] \\
$\omega_{b}$         & Uniform  & [0, 2$\pi$]  \\
\\
$P_{c}$             & Uniform  & [2, 50] \\
$T\rm{conj}_{c}$    & Uniform  & [2454000, 2456000]  \\
$e_{c}$               & Uniform  & [0, 1] \\
$\omega_{c}$              & Uniform  & [0, 2$\pi$] \\
\\
$K$  [km/s]                 & Uniform & [0.0,1.0] \\
\hline
\end{tabular}
\end{table}
\section{Details on the MCMC analysis}
\label{sec:MCMC}
For the S-index model using a GP, we used \textsc{RadVel}, which runs an MCMC analysis. We used the default values (such as the number of MCMC walkers= 50, 
number of steps=10000, and number of ensembles=8). For more details, readers can refer to \citet{Fulton18}. As stated in their paper, they used the Gelman-Rubin statistic for convergence by comparing the intra- and inter-chain variances, in which a value close to unity indicates the convergence of the chains, and checking them after every 50 steps. \\

For the Keplerian models (without activity indicators), we made use of \textsc{Pyaneti} which also runs an MCMC analysis. We also used the default  values (number of chains = 100, number of iterations =500, thin factor=10, and number of walkers=100) for these models. The convergence is reached when the Gelman criterion $\hat{R}$ is smaller than 1.02 for all the sampled parameters.
\newpage
\onecolumn

\setcounter{figure}{0}
\renewcommand{\thefigure}{B\arabic{figure}}

\begin{figure}
    \centering
    \includegraphics[width=1.0\textwidth]{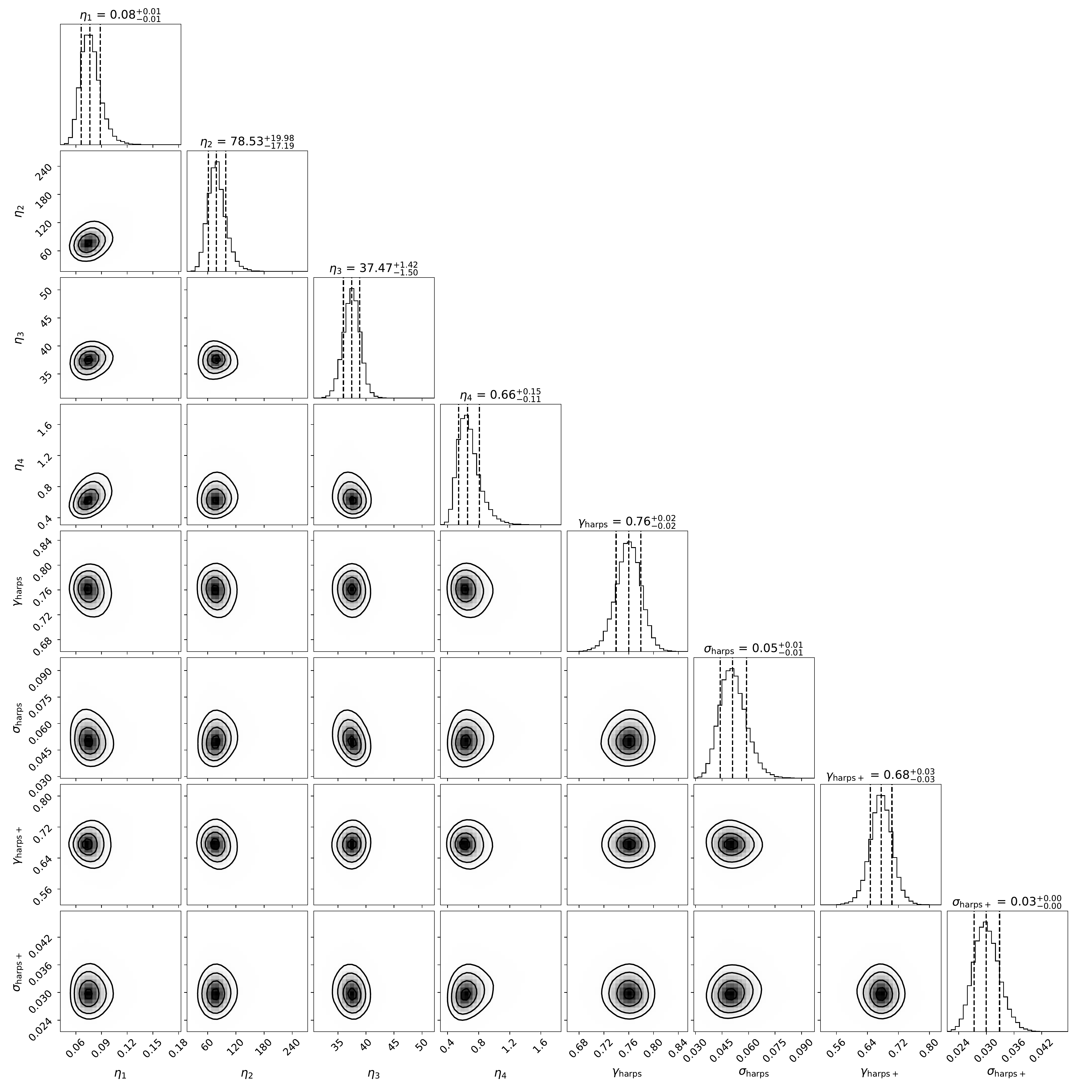}
    \caption{Marginalised posterior distributions for the parameters of the GP. The physical parameter $\eta_3$ indicates the stellar rotational period.}
    \label{fig:radvel-corner}
\end{figure}

\clearpage
\onecolumn
\section{Details on the RV and activity index models}
\label{sec:rv+act}

\begin{table}[tbh]
\caption{Summary of priors of the 1- and 2-Keplerian and activity index model (with GP).}
\label{tab:kep-act}
\centering
\begin{tabular}{c c c }
\hline\hline
Parameter& Prior& Range \\
\hline
$K_b$   [km/s]                & Uniform & [0.0,0.1] \\
$P_{b}$        & Uniform  & [3500, 4000] \\
$T\rm{conj}_{b}$     & Uniform  & [2456000, 24567000]  \\
$e_b\cos{\omega_b}$  & Uniform  & [-1, 1] \\
$e_b\sin{\omega_b}$  & Uniform  & [-1, 1] \\
\\
$K_c$  [km/s]                 & Uniform & [0.0,0.1] \\
$P_{c}$             & Uniform  & [2, 50]  \\
$T\rm{conj}_{c}$    & Uniform  & [2456000, 2457000]  \\
$e_c\cos{\omega_c}$  & Uniform  & [-1, 1] \\
$e_c\sin{\omega_c}$  & Uniform  & [-1, 1] \\
\\
GP$_\mathrm{rv,\eta_1}$ (h)& Uniform & [0.0,0.1]  \\
GP$_\mathrm{rv,\eta_4}$ (w)& Uniform & [0.0, 7.0] \\ 
GP$_{\eta_2}$ (lambda)&  Uniform & [0.0,1000]  \\
GP$_{\eta_3}$(theta) & Uniform &[10, 50]  \\
GP$_\mathrm{index,\eta_1}$ (h)&Uniform &[0.0, 2.0] \\
GP$_\mathrm{index,\eta_4}$ (w)&Uniform& [0.0, 7.0]\\
\\
HARPS-pre offset&Uniform & [13.300,13.400]  
 \\
HARPS-post offset& Uniform &[13.300,13.400] \\
HARPS-pre jitter& Uniform & [0.0,0.05]   \\
HARPS-post jitter &Uniform & [0.,0.05]   \\
index offset &Uniform &[0.0,1.0]  \\
index jitter &Uniform &[0.0,0.1]  \\
\hline
\end{tabular}
\end{table}
\section{Details on the best model (1-Keplerian and GP)}
\label{sec:best-model}

\begin{table}
\caption{Priors of the 1-Keplerian and GP model.}
\label{tab:1-kep-gp}
\centering
\begin{tabular}{c c c}
Parameter& Prior& Range \\
\hline\hline
$P_{b}$        & Uniform & [3500, 4200]  \\ 
$T\rm{conj}_{b}$ & Uniform & [2456200, 2456600] \\
$e_{b}$         & Uniform & [0.0, 0.1] \\ 
$\omega_{b}$      & Uniform  & [0, 2$\pi$]  \\
\\
$K $  [km/s]      & Uniform &  [0.0,0.02] \\ 
\\
GP$_{\eta_1}$ (A1)&  Uniform & [0.0,0.01]\\
GP$_{d\eta_1}$ (B1)& Uniform &[-0.002, 0.002] \\
GP$_\mathrm{\eta_2}$ ($\lambda_e$)&Uniform &[0.0, 2.0]  \\
GP$_\mathrm{\eta_3}$ (P$_{\mathrm{GP}}$)&Uniform &[20, 50] \\
GP$_\mathrm{\eta_4}$ ($\lambda_p$)&Uniform& [0.0, 7.0]\\
\hline
\end{tabular}
\end{table}

\setcounter{figure}{0}
\renewcommand{\thefigure}{F\arabic{figure}}

\begin{figure}
    \centering
    \includegraphics[width=1.0\textwidth]{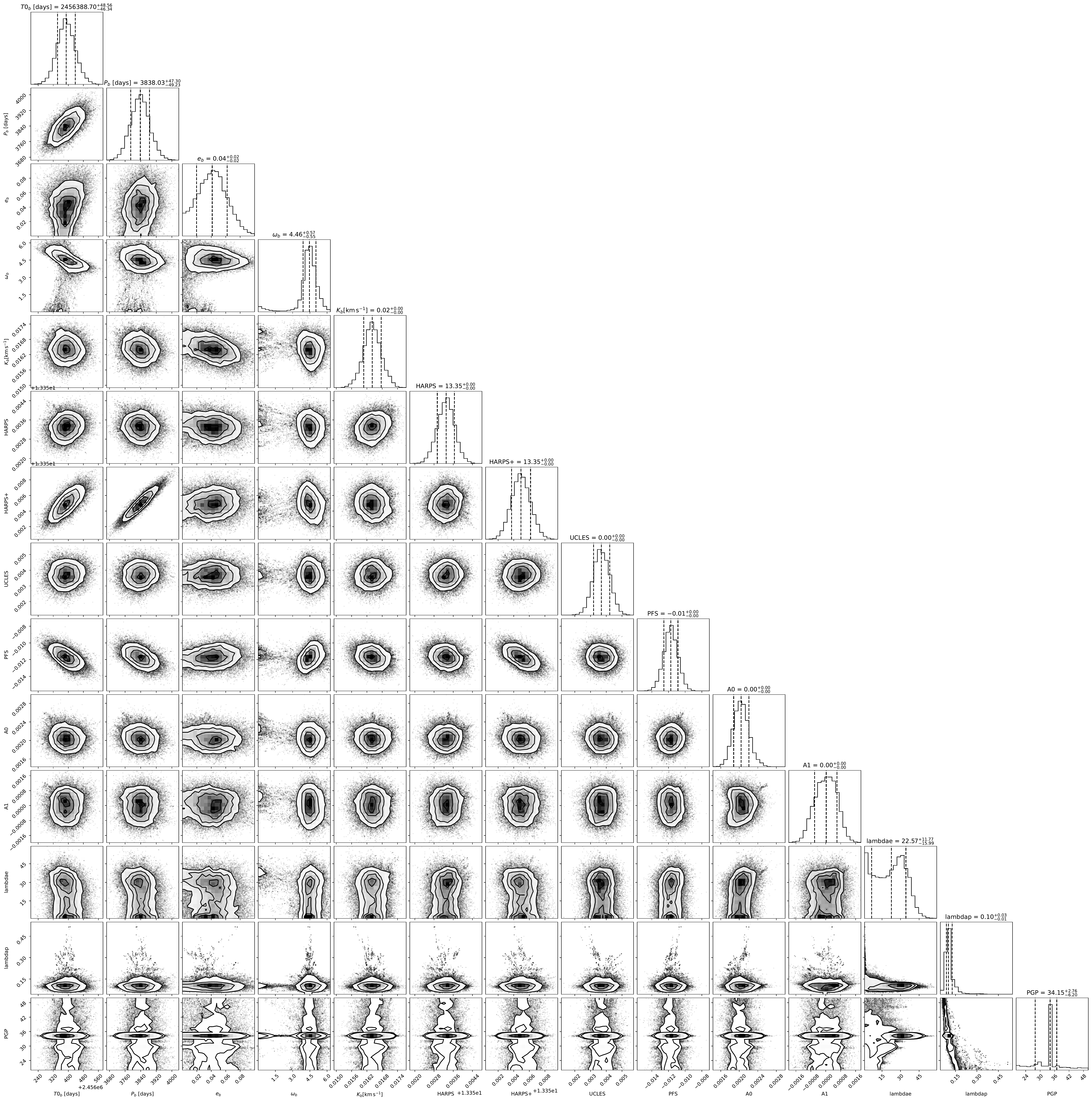}
    \caption{Corner plot from the 1-Keplerian and GP model.}
    \label{fig:corner-1k-gp}
\end{figure}

\end{appendix}

\end{document}